\documentclass[useAMS,usenatbib, onecolumn]{mn2e}
\usepackage{graphicx}

\title[Lower bounds of altitudes for pulsar $\gamma$-ray radiation]{Lower bounds 
of altitudes for pulsar $\gamma$-ray radiation}
\author[K.J.Lee, Y. J. Du, H. G. Wang, G. J. Qiao, R. X. Xu, and J. L. Han]{
K. J. Lee $^{1}$, Y. J. Du $^{3}$, H. G. Wang $^{2}$,
G. J. Qiao $^{1}$, R. X. Xu
   $^{1}$, J. L. Han $^{3}$\\
$^{1}${Astronomy Department, School of Physics, Peking University,
		 Beijing 100871, China (kjlee007@gmail.com)}\\
$^{2}${Center for astrophysics, Guangzhou University, Guangzhou 510400, China}\\
$^{3}${National Astronomical Observatories, Chinese Academy of Sciences, 20A 
Datun Road, Chaoyang, Beijing 100012, China}}

\begin{document}
\date{}
\pagerange{\pageref{firstpage}--\pageref{lastpage}} \pubyear{2009}
\maketitle
\label{firstpage}

\begin{abstract} Determining radiation location observationally plays a
very important role in testing the pulsar radiation models. One-photon
pair production in the strong magnetic field, $\gamma-e^{+}e^{1}$,
is one of the important physical processes in pulsar radiation
mechanisms. Photons near pulsar surface with sufficient energy will be
absorbed in the magnetosphere and the absorption optical depth for these
GeV $\gamma$-ray photons is usually large.  In this paper, we include the
aberrational, rotational and general relativistic effects and calculate
the $\gamma$-B optical depth for $\gamma$-ray photons.	Then we use the
derived optical depth to determine the radiation altitude lower bounds
for photons with given energies. As a case study, we calculate the lower
bounds of radiation altitudes of Crab pulsar for photons with energy
from 5 GeV to 1 TeV. \end{abstract}


\section{Introduction} 
{Pulsars are ideal laboratories to test physics
rules at extreme environment, yet how the pulsar radiations are
generated in pulsar magnetospheres is surely an open question though
more observational data are accumulated. Three types of models have been
proposed to explain the pulsed $\gamma$-ray radiation from pulsars. These
models are the polar cap model \citep{RS75, DH94}, the annular gap model
\citep{QLWXH03, QLZWX07}, and the outer gap model \citep{CHR86a, CHR86b,
CR94, CRZ00, ZFC07}.  The distinguishing characteristics of these models are
the acceleration region for primary particles. The primary acceleration
regions locate near the pulsar surface for polar cap models and for the
annular gap model \footnote{The annular gap model has extended radiation
region.}, but far away in the magnetosphere for outer. The classical
outer gap interacts with the annular gap due to the particle out flow
out from the annular gap, which behavior similar to the outer gap with
boundary currents \citep{Hirotani06}.  The fundamental physics
differs in three types of models due to the difference of magnetic field
intensity.  Observational judgments for the radiation altitudes are
therefore very critical to discriminate these radiation models. Early
observations indicated a 10 GeV cut-offs in the photon spectrum of
$\gamma$-ray pulsars \citep{Thom08}.  Thanks to the improvements of
$\gamma$-ray observational technologies, recent results both from
space-based Fermi Large Area Telescope (FERMI) and from ground-based
Cherenkov telescope (MAGIC) confirm such spectra cut-offs and also
have observed the high energy tails of the spectra cut-off above 10 GeV
\citep{MAG09, F09a, F09b}.  These existing high energy observations shied
lights on the altitudes of the pulsar $\gamma$-ray radiation regions.

The first order relativistic quantum processes allow interchange
of momentum between photons and magnetic field \citep{Erber66}. The
electron-positron pair production becomes no longer forbidden, such that a
$\gamma$-ray photon with sufficient energy can convert into $e^{+}e^{-}$
pairs in strong magnetic field. The cross section for the $\gamma$-B process 
increase with the background magnetic field and photon energy. Since the 
magnetic
field near pulsar surface is usually strong, the $\gamma$-B
absorption process is capable of developing a spectra cut-off in high energy 
ends of the pulsar spectra, if $\gamma$-ray photons are generated in the 
vicinity of star surface. 

After summarizing the works of \cite{BH01} and \cite{BHG97}, \cite{Bar04}
find an analytical formula to calculate such spectrum cut-off energy as a 
function of pulsar period, the pulsar surface magnetic field and the radiation
location altitude. One can fit the observed $\gamma$-ray spectra to
a theoretical model to measure the spectra cut-off energy, and then use
\cite{Bar04}'s formula to determine the lower bound for radiation
altitude. However in order to measure the spectrum cut-off energy,
one need to include both the absorption effects and the initial photon
source function \citep{BHG97}, which is sensitive to the input radiation
model. Naturally one could get different values for the cut-off
energy for the same data set by using different theoretical models. So
determining the radiation altitude lower bounds by using spectrum cut-off
energy is rather model-dependent, becase there is no model independent way of 
defining the value of cut-off energy.

In this paper, we present a model independent method to measure the
radiation altitude lower bounds. In fact, one can answer the question
about the lower bounds of radiation altitude through a slightly different
approach. Instead of using cut-off energy, one can ask what is the
lower bound for radiation altitude, such that $\gamma$-ray photons
with certain energy $\varepsilon$ will not be absorbed and can be
observed? Here, the term `not absorbed photons' and `observable photons' means 
those optical-thin photons \citep{RL86}, i.e.  photons with optical depth $\tau$ 
less than
a prescribed optical depth threshold $\tau_{\rm th}$, where $\tau_{\rm
th}$ is used to specify the confidence levels for low bounds of the radiation 
altitude. It turns out that there exists a minimal radius (called
\emph{last trapping radius}, $r_{\rm tr}$), below which the photons are
mostly absorbed and convert into $e^{+}e^{-}$ pairs in the presence of
intense magnetic fields. The existence of such last trapping radius is a
consequence of the $r^{-3}$ dependence of dipole magnetic field of pulsar,
i.e. we can find a small enough $r$ such that the magnetic field is
strong enough to initiate the $\gamma$-B absorption for the $\gamma$-ray
photons with sufficient energy. When pulsed $\gamma$-ray emission from a
pulsar is detected at a given energy band we can determine the $r_{\rm tr}$
correspondingly.  Obviously such $r_{\rm tr}$ is \emph{the lower bound
for radiation altitude for photons with certain energy}. Since such
lower bounds only involves the optical depth for absorption processes and 
observing energies, the $r_{\rm tr}$ is a model independent altitude lower bound.

In this paper, we dedicate to investigate the $\gamma$-B absorption
radiation transfer in the pulsar magnetosphere.  The $r_{\rm tr}$ is
calculated as a function of the magnetic field configuration of
pulsar magnetosphere and the photon energy. Analytical formula are
presented in \S~\ref{secest}. The more precise numerical calculations
are made in \S~\ref{secdet} for different source locations and in
\S~\ref{secradg} for phase resolved absorption. The application to Crab
pulsar with the phase resolved lower bounds for radiation altitudes are
presented in \S~\ref{secapp}.  Conclusions and discussions are given
in \S~\ref{seccon}.
}

\section{$\gamma$-B absorptions of high energy photons in pulsar magnetosphere}
\label{seccal}

\subsection{$\gamma$-B absorption---the analytical approach}
\label{secest}

In this section, we analytically estimate the optical depth and the radius for 
last trapping surface $r_{\rm tr}$. The $r_{\rm tr}$ is mathematically defined 
as the radius, at which the $\gamma$-B absorption optical depth is equal to a 
given threshold optical depth $\tau_{\rm th}$.  The probability is less than 
$e^{-\tau_{\rm th}}$ for photons to escape from a region with altitude lower 
than $r_{\rm tr}$. For examples, when $\tau_{\rm th}=1, 10$, the escape 
probabilities are $37\%$ and $5\times 10^{-5}$ respectively \citep{RL86}.  
Obviously, such probability tells about the confidence of $r_{\rm th}$ as an 
altitude lower bound,

Here we only consider magnetic field much weaker than $10^{13}$ Gauss, and the
photon energy is much larger than $e^{+}e^{-}$ static mass, i.e. much larger
than 1 MeV. In this way, we can ignore the photon splitting effect \citep{BH01}.
Meanwhile the number of pair states is large enough to allow us ignore the
resonance effects due to the finite number of pair states \citep{DH83}. The
following formula is appropriate to calculate the $\gamma$-B absorption
coefficient $\kappa$ \citep{Erber66}

\begin{equation}
\kappa=1.55 \times 10^7 \varepsilon_{\rm [MeV]}^{-1} K_{1/3}^2 \left(
\frac{30.0}{B_{\bot [12]}
\varepsilon_{\rm [MeV]}}\right) {\rm cm^{-1}}.
\label{eq:absor}
\end{equation}
The function $K_{1/3}$ is the second
type modified Bessel function of order $1/3$; $B_{\bot [12]}$ is the magnetic
field strength perpendicular to the photon propagating direction in unit of
$10^{12}$ Gauss; and $\varepsilon_{\rm [MeV]}$ is the energy of photon in unit of 
MeV.  

The photons of 5 GeV to 1 TeV in the pulsar magnetosphere are radiated by 
ultra-relativistic charged particles, which are approximately beamed along the 
magnetic
field lines. After a photon propagates
a length of $\lambda$, the angle $\theta_{\rm i}$ between photon direction and 
local
magnetic field becomes $\theta_{\rm i}\sim \lambda/r_{\rm cur}$. Here $r_{\rm 
cur}$
is the curvature radius of magnetic field lines.  For dipole magnetic field,  we
have $r_{\rm cur}\sim \frac{4}{3}r^{1/2} r_{\rm lc}^{1/2}$, where $r$ is the 
altitude of the photon and $r_{\rm lc}$ is the light cylinder radius.  Using 
geometrical unit such that light velocity is 1, we have $r_{\rm lc}\sim 
p/(2\pi)$ for the pulsar with period of $p$. One gets $B_{\rm
\bot}\sim \theta_{\rm i} B=\lambda/r_{\rm cur} B$, where $B$ is the magnetic 
field intensity at
the altitude. The dipole magnetic field has intensity
of $B\sim B_{0} R^3/r^3$, with $B_{0}$ as the magnetic field
intensity at the pulsar surface, and $R$ as the radius of the pulsar. The 
optical depth $\tau=\int_{0}^{\infty} \kappa d\lambda$. Because the 
absorption coefficient $\kappa$ exponentially dependent on the magnetic 
field, which is of $r^{-3}$ dependence, the absorption coefficient exponentially
decrease as $e^{-\lambda^3/\theta_{\rm i}}$, when $\lambda$ is large. The
dominant part of the optical depth are thus the integration of $\lambda$ from 0
to $r_{\rm s}$, where $r_{\rm s}$ is the altitude, at which the photons are
generated (see the Appendix for the details). Further more, when $\lambda \le
r_{\rm s}$, the magnetic field $B$ can be approximately regarded as a constant.
Thus the \emph{characteristic absorption length} $\lambda_{\rm c}$ is
$\lambda_{\rm c}\simeq r_{\rm s} $, and the integration for optical depth is
then replaced by $\tau\sim \kappa \lambda_{\rm c }\sim \kappa r_{\rm s}$, which
is later further justified by numerical integration. Thus \begin{equation}
\tau(r_{\rm s})\sim \frac{{1.55 \times 10^7 r_{[{\rm{cm}}]} }}{{\varepsilon
_{[{\rm{MeV}}]} }}K^2 _{1/3} \left( {\frac{{2.76 \times 10^6 r_{s, [{\rm{cm}}]}
^{5/2} p_{[s]} ^{1/2} }}{{B_{0,[12]} R_{[{\rm{cm}}]} ^3 \varepsilon
_{[{\rm{MeV}}]} }}} \right), \label{eqtauest} \end{equation} where $r_{s, \rm
[cm]}$ is the $r_{\rm s}$ in unit of cm, $p_{\rm [s]}$ is the period of a pulsar
in unit of second.

\begin{figure} \center \includegraphics[height=2.0 in]{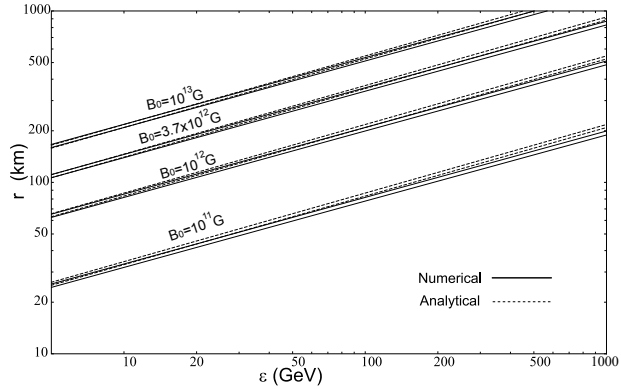}
	\caption{The radius of last trapping surface $r_{\rm tr}$ for
	different photon energy $\varepsilon$ and surface magnetic
	field.The x-axis is the photon energy, while the star surface magnetic field 
	intensity $B_0$ are labeled above each curve. For each
	$B_{0}$ there are two groups of curves.  Two solid curves
	are the numerical result from directly solving $\tau(r_{\rm
	tr})=\tau_{\rm th}$ with $\tau_{\rm th}=1,10$ from the top
	down. Three dashed curves are
	analytical solutions from Eq.~\ref{eq:eqtr} with $\tau_{\rm
	th}=1,10$.}
\label{fig:estibond} \end{figure}

By setting $\tau(r_{\rm tr})=\tau_{\rm th}$ and solving for $r_{\rm tr}$, we get 
radius of last trapping surface, which should be the
lower bounds for the radiation altitude for photons.  The numerical
solutions to $\tau(r_{\rm tr})=\tau_{\rm th}$ is given in 
Fig.~\ref{fig:estibond}, for different $\tau_{\rm th}$.

We can also get analytical solutions to $\tau(r_{\rm tr})=\tau_{\rm th}$ using 
asymptotic method. Note that the Bessel function have
asymptotic property of $K_{1/3}^2(x)\simeq \pi e^{-2x} /(2x)$. The solution to
Eq.~\ref{eqtauest}  can be presented by using the Lambert $W$ function, where
function $y=W(x)$ is defined as the solution to equation $y e^{y}=x$
\citep{CGH96}.  Using the function $W$ and taking $R= 10^6$ cm, the solution to
$\tau(r_{\rm tr})=\tau_{\rm th}$ is \begin{equation} r_{\rm tr, [cm]}= 2.6
\times 10^4 B^{2/5} _{0,[12]} \varepsilon ^{2/5} _{[{\rm{MeV}}]} p_{[{\rm{s}}]}
^{ - 1/5} W^{2/5} \left( {\frac{{3.46 \times 10^{20} B^{2/3} _{0,[12]}
}}{{p_{[{\rm{s}}]} ^{1/3} \varepsilon _{[{\rm{MeV}}]} } \tau_{\rm th}^{5/3}}}
\right) \end{equation} which have asymptotic approximation given as
\begin{equation} r_{\rm tr, [cm]}\simeq(1.1 \times 10^5 -1.9\times 10^{3}
\ln\tau_{\rm th})  B_{0, [12]}^{2/5} \varepsilon_{\rm [MeV]}^{2/5}p_{\rm
[s]}^{-1/5}.  \label{eq:eqtr} \end{equation} This analytical results are plotted
together with the numerical results in Fig.~\ref{fig:estibond}. The analytical
approximation give error of about 20\% percent. {The effect of $\tau_{\rm th}$ 
is
very tiny, due to the logarithmic dependence of $\tau_{\rm th}$ as we expect 
from Eq.~\ref{eq:eqtr}. We present the results with $\tau_{\rm th}=1, 10$. It 
turns out that the difference between the $r_{\rm tr}$ for
$\tau_{\rm th}=1$ and for $\tau_{\rm th}=100$ is still less than 20\%, where the
probability for photons escaping region with $\tau_{\rm th}=100$ is $10^{-43}$ 
times smaller than escaping region with $\tau_{\rm th}=1$, or the radiation from  
region with
$\tau_{\rm th}=100$ is $10^{-43}$ timeas fainter than radiation from $\tau_{\rm  
th}=1$.  This is far beyond dynamical range of instruments and is exactly why we 
can define the optical thin photons as
observable photons and use $r_{\rm tr}$ as the altitude low bound, without going 
into the details of equipment response.}

For the case of Crab pulsar, i.e.  $B_{0, \rm [12]}=3.7$, $p_{\rm [s]}=0.033$
\citep{MHT05}, we see from Fig.~\ref{fig:estibond} that the radius of the last 
trapping surface for 5 GeV, 10 GeV,
100 GeV  and 1 TeV $\gamma$-ray radiation are 100 km, 150 km, 350 km, 850
km respectively. 

\subsection{$\gamma$-B absorption processes with detailed geometry}
\label{secdet}

We have estimated the radius of last trapping
surface $r_{\rm tr}$ and taken it as lower bounds of the radiation altitudes for 
the pulsed pulsar $\gamma$-ray radiation. The $\gamma$-B process dependents on 
geometries
of both photon ray directions and magnetic fields. In this section,
we calculate the $r_{\rm tr}$ with detailed geometry.

The $\gamma$-B absorption coefficient $\kappa$ is sensitive to the
geometrical parameter $\theta_{\rm i}$, the angle between the photon
propagating direction and the local magnetic field direction. There
are five possible geometrical effects that changes the $\theta_{\rm
i}$. These effects are listed as follows. (1).  Photons, when generated,
are not exactly beamed along local magnetic field lines due to the finite
Lorentz factor $\gamma_{\rm p}$ of particles.  This introduces an increase
of $\theta_{\rm i}$ by $\delta \theta_{\rm i}$, where $\delta\theta_{\rm
i}\sim \gamma_{\rm p}^{-1}< 10^{-2}$ given $\gamma_{\rm p}\gg10^{2}$
\citep{RL86}. (2). Due to special relativistic effect, the rotational
velocity of the pulsar boosts the photons to a new direction, which gives
$\delta\theta_{\rm i}\sim \Omega r\sim r/r_{\rm lc}\sim 10^{-2}$ for a
young pulsar with $p\sim 0.1$ s \citep{Gad05}, where $\Omega=2\pi/p$ is
the pulsar's rotational angular velocity. The $r$ is choose to be 100
km as order of magnitude estimation from previous section. It should be
noted that we use geometrical unit through out this paper without special
mention, i.e. the light speed $c=1$ and the gravitational constant $G=1$.
(3). The magnetic field is co-rotating with the star, thus after the
photon propagates for time of $t$, the magnetic field has rotated for an
angle of $\Omega t$. It gives $\delta\theta_{\rm i} \sim \Omega t \sim
2\pi r/p\sim 10^{-2}$. (4). Due to the gravitational field of pulsar, the
photons propagate along `curved' geodesics rather than `straight'
lines. This effect is order of $\delta\theta_{i}\sim\theta_{i} m/r\sim
10^{-3}$, as $m\sim2$ km for 1.4 solar mass pulsar.  (5). The pulsar drags the 
background space co-rotating with it due to gravitomagnetism
effects. The space co-rotating angular velocity is Lense-Thirring
angular velocity $\omega_{\rm LT}\sim 2 m \Omega R^2/r^3$, where $R$ is
the pulsar radius.  This gives $\delta\theta_{\rm i}\sim\omega_{\rm LT}
t \sim 2 m \Omega R^2/r^2\sim 10^{-7}$.  We see that the curved spacetime
effect and frame-dragging effects are of higher order geometrical effect
compared to other effects, which are coincident with the results of
\cite{GH94}. Therefore, we only consider the aberration effect and
the magnetic field rotation effect to correct the geometrical effects
to the order of $10^{-2}$. This conclusion are valid for short period
pulsar ($p\ge$0.1s) and higher altitude absorption ($r\ge$100km).  Thus,
we can just use flat space-time geometry (including photon direction
and magnetic field direction) to calculate the geometrical parameters
(e.g. $\theta_{\rm i}$).

However, the gravitational effects
also play the other two important roles in $\gamma$-B processes.  (1). The 
gravitational field makes the photon red shifted, i.e.  the energy $\varepsilon$
of photons observed by static observer is $\varepsilon=\varepsilon_{\infty}
(1+\eta)$, where $\varepsilon_{\infty}$ is the photon energy observed by
infinite-distant observer and $\eta=m/r$ is the Newtonian gravitational 
potential. This introduces energy correction by $\varepsilon m/r\sim 
10^{-2}\varepsilon$.
(2). The dipole magnetic field strength is enhanced due to gravitational effect 
\citep{GO66, AC70, WS83, MT86}, which magnetic field intensity correction of 
order $m/r\sim 10^{-2}$.
Both of the two effects are of
$10^{-2}$ order effects. In short,
we can use flat space geometry to calculate the geometrical parameters, i.e.  
$\theta_{i}$, but the magnetic field and photon energy need the first order 
gravitational corrections to include effects of $10^{-2}$ level.

\begin{figure} \center \includegraphics[height=2.0 in]{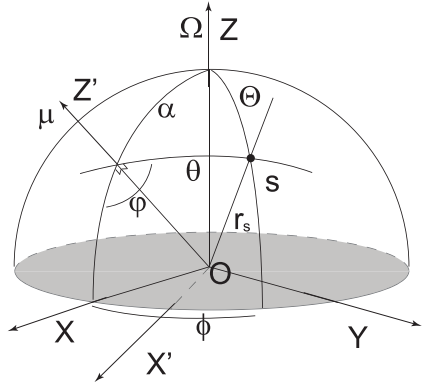}
	\caption{The geometrical configuration for the coordinate systems for 
	calculations. In the laboratory frame $O-XYZ$,
the $Z$ axis is aligned with the rotational axis $\rm \bf \Omega$ of
pulsar. The magnetic frame $O-X'YZ'$ is generated by rotating the $O-XYZ$
coordinate around $Y$ axis by an inclination angle of $\alpha$, such that the 
$Z'$ axis is aligned with
the dipole magnetic momentum $\rm \bf \mu$. The polar angle and azimuthal angle 
in laboratory
polar coordinate are denote as $\Theta, \Phi$, while the polar angle and
azimuthal angle in the magnetic polar coordinate are $\theta,\varphi$. The 
$r_{\rm s}$ is the radius for the radiation source.}
\label{geoconf}
\end{figure}

We set up the
coordinates as shown in Fig.~\ref{geoconf}. In the coordinate $O-XYZ$,
the $Z$ axis is aligned with the rotational axis $\rm \bf \Omega$ of
pulsar. This coordinate $O-XYZ$ is called the laboratory frame. Another
coordinate, the $O-X'YZ'$, is generated by rotating the $O-XYZ$
coordinate around $Y$ axis by inclination angle $\alpha$, such that the $Z'$ 
axis is aligned with
the dipole magnetic momentum $\rm \bf \mu$. We call
this coordinate the magnetic frame. The two vectors $\rm
\bf \Omega$ and $\rm \bf \mu$ locates in the plane $O-XZ$, which is
called $\Omega-\mu$ plane. The polar coordinate associated with $O-XYZ$
and $O-X'YZ'$ are called the laboratory polar coordinate and the magnetic
polar coordinate, respectively. The polar angle and azimuthal angle
in the laboratory polar coordinate are denoted as $\Theta, \Phi$, while
the polar angle and azimuthal angle in the magnetic polar coordinate
as $\theta,\varphi$. Here, we use bold type font to label the vector or the
matrix; while we use subscript $_{x,y,z}$ and $_{x', y', z'}$ to indicate
their component value in the laboratory frame and component value in the 
magnetic
coordinate respectively. The coordinate transformation between
two coordinates is given by $\rm \bf a^{i}=T^{ij}\cdot a'^{j}$, where
$\rm \bf a$ and $\rm \bf a'$ are any vectors in laboratory frame and
magnetic frame respectively, transformation $\rm \bf T_\alpha^{ij}$
is the matrix given as

\begin{equation}
	{\rm \bf 	T_\alpha^{ij}}=
\left(
\begin{array}{lll}
 \cos \alpha & 0 & \sin \alpha  \\
 0 & 1 & 0 \\
- \sin \alpha  & 0 & \cos \alpha 
\end{array}
\right)
\label{codtrans1}
\end{equation}

Let's consider a photon generated at the position 'S' in Fig.~\ref{geoconf}.  
The initial position of photon $\rm \bf r_{\rm s}$ in the magnetic polar 
coordinate is 

\begin{equation}
\left( {\begin{array}{*{20}c}
	{r_{\rm s, x'} }  \\
   {r_{\rm s, y'}}  \\
   {r_{\rm s, z'}}  \\
\end{array}} \right) = r_{s}\left( {\begin{array}{*{20}c}
	{\sin \theta_{\rm s} \cos \varphi_{\rm s} }  \\
   {\sin \theta_{\rm s} \sin \varphi_{\rm s} }  \\
   {\cos \theta_{\rm s} }  \\
\end{array}} \right),
\end{equation}
where $\theta_{\rm s}, \varphi_{\rm s}$ are the polar angle and azimuthal angle 
for source position `S' in magnetic frame, and $r_{\rm s}$ is the radiation 
altitude.  The photon generated at position `S' propagates along direction $\rm 
\bf n_{\rm B}$ of the local magnetic field for a co-rotating observer.  The 
component value of vector $\rm \bf n_B$ is easily calculated in magnetic polar 
coordinates \citep{SW06}

\[
\left( {\begin{array}{*{20}c}
	{n_{\rm B, x'} }  \\
   {n_{\rm B, y'}}  \\
   {n_{\rm B, z'}}  
\end{array}} \right) = \left( {\begin{array}{*{20}c}
   {\frac{{3\cos \theta_{\rm s} \cos \varphi_{\rm s} \sin \theta_{\rm s} }}{{\sqrt {1 + 3\cos ^2 \theta_{\rm s} } }}}  \\
   {\frac{{3\cos \theta_{\rm s} \sin \varphi_{\rm s} \sin \theta_{\rm s} }}{{\sqrt {1 + 3\cos ^2 \theta_{\rm s} } }}}  \\
   {\frac{{1 + 3\cos 2\theta_{\rm s} }}{{\sqrt {1 + 3\cos ^2 \theta_{\rm s} } }}}  
\end{array}} \right),
\]
which can be used to calculate the component value of $\rm \bf n_B$ in 
laboratory frame by combining it with Eq.~\ref{codtrans1}. 

The rotational effect boosts the photon to direction $\rm \bf n_{\nu}$, known as 
the aberration effect.
The relation between components of $\rm \bf n_{\nu}$ and components $\rm \bf 
n_{\rm B}$ is give by Lorentz transformation \citep{MTW73}

\[
\left( {\begin{array}{*{20}c}
   \chi   \\
   {\chi n_{\rm \nu, x} }  \\
   {\chi n_{\rm \nu, y} }  \\
   {\chi n_{\rm \nu, z} }  \\
 \end{array}} \right) = {\cal L}\cdot\left( {\begin{array}{*{20}c}
   1  \\
   {n_{\rm B, x} }  \\
   {n_{\rm B, y} }  \\
   {n_{\rm B, z} }  \\
\end{array}} \right),
\]

where matrix $\cal L$ is defined as
\[
{\cal L}=\left( {\begin{array}{*{20}c}
   \gamma  & {\gamma v_x } & {\gamma v_y } & 0  \\
   {\gamma v_x } & {1 + (\gamma  - 1)\frac{{v_x ^2 }}{{\beta ^2 }}} & {(\gamma  - 1)\frac{{v_x v_y }}{{\beta ^2 }}} & 0  \\
   {\gamma v_y } & {(\gamma  - 1)\frac{{v_x v_y }}{{\beta ^2 }}} & {1 + (\gamma  - 1)\frac{{v_y ^2 }}{{\beta ^2 }}} & 0  \\
   0 & 0 & 0 & 1  \\
\end{array}} \right),
\]
and $\gamma$, $\beta$ are $\gamma=1/\sqrt{1-v_x^2-v_y^2}$ and 
$\beta=\sqrt{v_x^2+v_y^2}$ respectively, $\chi$ is the redshift of photon due to 
aberration effect.
The $v_x, v_y$ is the components of co-rotation velocity at the place where 
photon is generated. The velocity can be calculated  by $\rm \bf v=\Omega \times 
r_{s}$.  In the laboratory frame, we have

\[
\left( {\begin{array}{*{20}c}
	{v_{\rm x} }  \\
	{v_{\rm y} }  \\
	{v_{\rm z} }  \\
\end{array}} \right) = r_{\rm s}\Omega \left( {\begin{array}{*{20}c}
   { - \sin \theta_{\rm s} \sin \varphi_{\rm s} }  \\
   {\cos \theta_{\rm s} \sin \alpha  + \cos \alpha \cos \varphi_{\rm s} \sin \theta_{\rm s} }  \\
   0  \\
\end{array}} \right)
\]

After take such aberration effect into account, the equation for component value 
for ${\rm \bf n}_{\rm \nu}$ is

\begin{equation}
\left \{ \begin{array}{c}
n_{\nu,x} = \frac{3 h+f \sin \alpha -r_{\rm s} \Omega  \sin \theta_{\rm s}  \sin 
\varphi_{\rm s} [\sqrt{2} b+3 d r_{\rm s} \Omega  \sin (2\theta_{\rm s})
 \sin \varphi_{\rm s} ]}{\sqrt{2} g b \chi} \\
n_{\nu, y} = \frac{2 b d r_{\rm s} \Omega +\sqrt{2} [6 \cos \theta_{\rm s} -d 
r_{\rm s}^2 \Omega ^2 (3 h+f \sin \alpha )] \sin \theta_{\rm s}  \sin
	\varphi_{\rm s} }{2 g b\chi}\\
n_{\nu,z}=\frac{f \cos \alpha -3 h \tan \alpha }{\sqrt{2} b\chi}\\
\end{array} \right.
\label{eqnxnynz}
\end{equation}

where 

\[
\begin{array}{c}
	g=\sqrt{1-r_{\rm s}^2 \Omega ^2 \left[(\cos \alpha \sin\theta_{\rm s} \cos 
	\varphi_{\rm s}+\sin\alpha \cos \theta_{\rm s} )^2+\sin ^2\theta_{\rm s}  
	\sin
	^2\varphi_{\rm s} \right]},\\
   b=\sqrt{3 \cos (2 \theta_{\rm s} )+5}, \\
   d=\cos \alpha \sin\theta \cos \varphi_{\rm s} +\sin\alpha \cos \theta_{\rm 
   s}\\
   f=3 \cos (2 \theta_{\rm s} )+1, \\
   h=\cos \alpha \sin (2 \theta_{\rm s} ) \cos\varphi_{\rm s}, \\
   \chi=\frac{b+\sqrt{2} r_{\rm s} \Omega  \sin \alpha  \sin \theta_{\rm s}  
   \sin \varphi_{\rm s}  }{g b} .  \end{array}
\]

The position of photon ${\rm \bf x}(t)$ is a function
of photon propagating
time $t$. With the ${\rm \bf n}_{\rm \nu}$, we can calculate the position of  
photon by
${\rm \bf x}(t)={\rm \bf r}_{\rm s}+{\rm \bf n_{\nu}} t$.  We need to further 
calculate the magnetic field at each position ${\rm \bf x}(t)$ to get the 
absorption coefficient $\kappa$.  To do this, we firstly calculate magnetic 
field in the magnetic polar coordinate and then use coordinate transformation to 
derive the magnetic field strength at $\rm \bf x$.

The magnetic field in the co-rotating magnetic polar coordinate is

\begin{equation}
	\label{eqbint}
{\rm \bf B'} = \frac{{B_0 R^3 }}{{2r^3 }}\left( {1 + \eta \frac{{8 + 4\cos 
(2\theta )}}{{5 + 3\cos (2\theta )}}} \right)\sqrt {\frac{{5 + 3\cos (2\theta 
)}}{2}} {\rm \bf n}_{\rm B}
\end{equation}
where $r, \theta, \varphi$ are the radius, polar, and azimuthal angle in the 
co-rotating magnetic polar coordinate for ${\rm \bf x}(t)$, $B_0$ is the 
effective surface magnetic field determined from energy losing rate, and 
$\eta=m/r$ is the Newtonian gravitational potential. The magnetic field 
configuration defined in Eq.~\ref{eqbint} satisfy two conditions that the 
magnetic field intensity is concord with results in the Schwarzschild background 
and the magnetic field direction is concord with the flat background results.
Due to the rotation of pulsar, the magnetic field is also `rotating'.  Then the 
relation between magnetic field in laboratory frame and the magnetic field in 
the co-rotation magnetic polar coordinates is $\rm \bf B=T_\Omega \cdot 
T_\alpha\cdot B'$, where the transformation $T_\Omega$ is coordinates 
transformation due to the rotation of the pulsar. The matrix form of $\rm \bf 
T_{\Omega}$ is given by

\[
{{\rm \bf T}_{\rm \Omega}}   = \left( {\begin{array}{*{20}c}
   {\cos \Omega t} & { - \sin \Omega t} & 0  \\
   {\sin \Omega t} & {\cos \Omega t} & 0  \\
   0 & 0 & 1  \\
\end{array}} \right). \]
Use this transformation, we can also figure out the relation between the
co-rotational magnetic polar coordinates $\{r, \theta, \varphi\}$ and laboratory 
frame $\{x,y,z\}$, which is given as

\begin{equation}
\left( {\begin{array}{*{20}c}
   x  \\
   y  \\
   z  \\
\end{array}} \right) = r{{\rm \bf T}_{\rm \Omega}}\cdot \left( 
{\begin{array}{*{20}c}
  {\cos \theta \sin \alpha  + \cos \alpha \cos \varphi \sin \theta }\\
   {\sin \theta \sin \varphi}  \\
   {\cos \alpha \cos \theta  - \cos \varphi \sin \alpha \sin \theta }  \\
\end{array}} \right).
\label{eqphotpos}
\end{equation}
Since the photo propagating direction does not change in laboratory frame, the 
perpendicular magnetic field strength $B_{\bot}$ is $B_{\bot}=\sqrt{\rm \bf 
B\cdot B-(B\cdot n_{\nu})^2}$. When the gravitational redshift of photon is 
taken into account, we have $\varepsilon=\varepsilon_{\infty} (1+\eta)$.  In 
this way, the absorption coefficient $\kappa$ as a function of $t$ can be 
calculated using the $\varepsilon$, $B_{\bot}$, and Eq.~\ref{eq:absor}. We then 
integrate $\kappa$ over $t$ to calculate the optical depth.

We now summarize the steps to calculate the optical depth and the radius of last 
trapping surface, as follows,

\begin{enumerate}
	\item For a radiation source located at $\{r_{\rm s}, \theta_{\rm s}, 
		\varphi_{\rm s} \}$, we calculate the photon propagating direction ${\rm 
		\bf n}_{\nu}$ with Eq.~\ref{eqnxnynz};
	\item We calculate the photon position after time $t$ using ${\rm \bf 
		x}(t)={\rm \bf r}_{\rm s}+{\rm \bf n_{\nu}} t$;
	\item Solve Eq.~\ref{eqphotpos} to calculate the coordinate of photon in 
		magnetic polar coordinate ($r,\theta,\varphi$ ) from coordinate of ${\rm 
		\bf x}(t)$;
	\item Using the photon position($r,\theta,\varphi$ ) to calculate the 
		magnetic field strength according to Eq.~\ref{eqbint}.  Then calculate 
		$B_{\bot}$ using $B_{\bot}=\sqrt{\rm \bf B\cdot B-(B\cdot n_{\nu})^2}$;
	\item Using $\tau(r_{\rm s})=\int_{0}^{\infty} \kappa(t; r_{\rm s}) dt$ to 
		calculate the optical depth for the photon coming from position $\rm \bf 
		r_{\rm s}$.  The $\kappa$ is calculated from Eq.~\ref{eq:absor};
	\item Solve $\tau(r_{\rm tr})=\tau_{\rm th}$ respected to $r_{\rm tr}$ to 
		determine the lower bounds for radiation altitude given required 
		$\tau_{\rm th}$;
\end{enumerate}

\begin{figure*} \center \includegraphics[height=4.0 in]{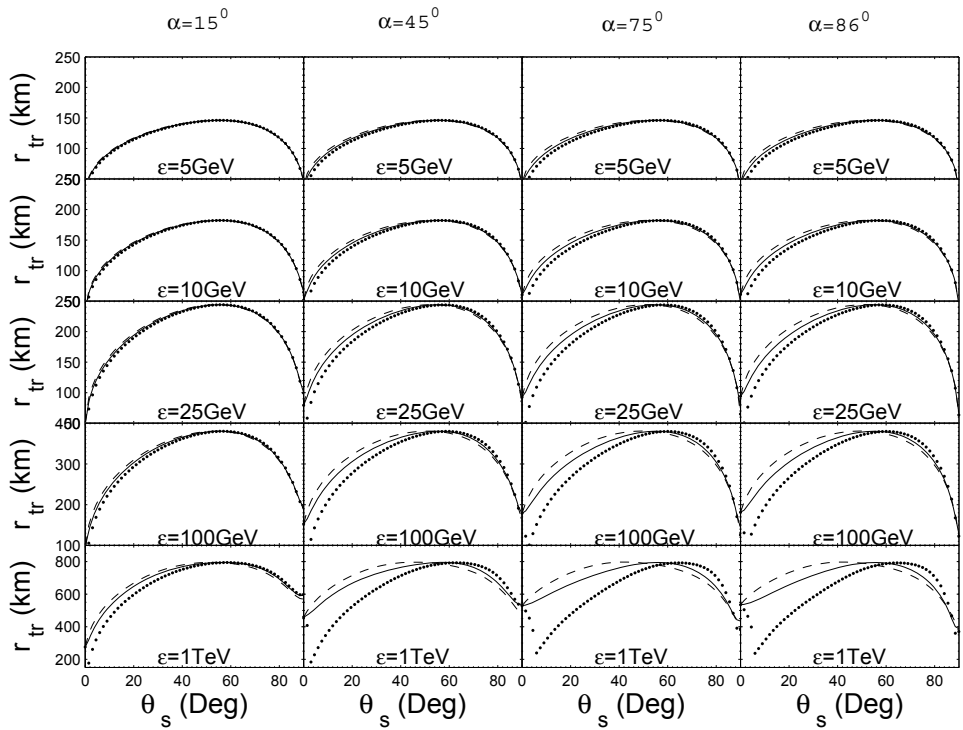}
	\caption{The lower bounds $r_{\rm tr}$ of altitude as functions of radiation 
	source polar angle $\theta_{\rm s}$.
	The solid lines, dashed lines and dotted lines correspond to $\varphi_{\rm 
	s}=0^{o}$, $\varphi_{\rm s}=90^{o}$, and $\varphi_{\rm s}= 270^{o}$ 
	respectively. The surface magnetic field strength is taken to be 
	$B_0=3.7\times 10^{12}$ Gauss as the Crab pulsar, and the threshold optical 
	depth is set to $\tau_{\rm th}=1$. The $\varepsilon$ is the photon energy 
	used in calculations for each panel, while the inclination angle of pulsar 
	$\alpha$ is labeled on the top of each column. The results for $\tau_{\rm 
	th}=3$ and $\tau_{\rm th}=10$ are almost the same as results presented here.  
	}
\label{fig:bondvstp1}
\end{figure*}

There is no analytical solution to $\tau(r_{\rm tr})=1$ respected $r_{\rm tr}$.  
We use bi-section method to solve it numerically, while the integration is 
performed using adaptive integration method to refine a preset logarithmic mesh 
of $t$ to achieve necessary numerical precision. The results are given in 
Fig.~\ref{fig:bondvstp1}. We also get the results for $\tau_{\rm th}=3$ and 
$\tau_{\rm th}=10$, the plots are very similar to Fig.~\ref{fig:bondvstp1} due 
to the logarithmic dependence of $\tau_{\rm th}$.

The results are different for various inclination angles of pulsars, due to the 
aberration effect and rotational magnetic field effect. If these two effects 
would be
ignored, the results would be insensitive to inclination angle $\alpha$. We also
see the aberration effects become important for large inclination angles, since 
the $r_{\rm s}$ shows larger difference between $\varphi_{\rm s}=90^{o}$ (for 
the leading photon rays) and $\varphi_{\rm s}=270^{o}$ (for the trailing photon 
rays) cases.

\subsection{Radiation Geometry and Phase-resolved Lower Bounds for Radiaiton 
Altitudes}
\label{secradg}

In this section we determine the radiation location for different longitudinal 
phase.  The details of radiation geometry can be found in \cite{GGR84, LM88, 
LQW06}.  We omit the aberration effects here, because it is second order effect 
for calculating the pulse phase.

Given the pulse profile longitude $\Delta\Phi$ (see Fig.~\ref{geoconf} for 
details) and the view angle $\zeta$, the half angular beam width for the 
radiation beam $\theta_{\mu}$ can be solved by\citep{GGR84, LM88} 

\begin{equation}
	\sin^2\left(\frac{\theta_{\mu}}{2}\right)=\sin^2\left(\frac{\Delta\phi}{2}\right)\sin\alpha\sin\zeta+\sin^2\left(\frac{\zeta-\alpha}{2}\right),
\label{rho} \end{equation}

With the $\theta_{\mu}$, the radiation source polar angle $\theta_{\rm s}$ and 
azimuthal angle $\varphi_{\rm s}$ in magnetic polar coordinate can be 
calculated, where
\begin{equation}
	\theta_{\rm s}=\frac{1}{2} \arccos \left[\frac{\sqrt{\sin ^4\theta_\mu -10 
	\sin
^2\theta_\mu +9}-\sin ^2\theta_\mu}{3} \right],
\label{eqtheta}
\end{equation}
and
\begin{equation}
	\label{eqvarphi}
	\varphi_{\rm 
	s}=\arccos\left[\frac{\cos\alpha\cos\theta_{\mu}-\cos\zeta}{\sin\alpha\sin\theta_{\mu}}\right  
	]
	.
\end{equation}

Using Eq.~\ref{rho}, \ref{eqtheta}, and \ref{eqvarphi}, we can calculate the 
angular position for radiation source $\theta_{\rm s}, \varphi_{\rm s}$ from 
pulse profile phase $\Delta\Phi$ and view angle $\zeta$. The we can determined 
the $r_{\rm tr}$ using the techniques developed in \S~\ref{secdet}.  For 
$\zeta=60^{\circ}$ case, the result could be found in Fig.~\ref{fig:phra}.

\subsection{Application to Crab pulsar}
\label{secapp}
As a case study, we apply the method for altitude lower bounds to Crab pulsar 
and obtain
the phase resolved altitude lower bounds for 25 GeV $\gamma$-ray emission. The 
view angle of Crab pulsar is taken to be $60^0$, as measured from X-ray image of 
Crab nebular \citep{WHT00, Wang03} and $\alpha\sim 55^{\circ}-60^{\circ}$ by 
fitting slot gap model or $\alpha\sim70^{\circ}$ from fitting outer gap model 
\citep{F09b}.  The surface magnetic field of Crab is taken to be 
$B_{0}=3.7\times 10^{12}$ Gauss.  The inclination angle for Crab is unknown at 
this time, thus we calculate the $r_{\rm tr}$ for the four cases, $\alpha=15^o, 
45^o, 75^o, 86^{o}$ \footnote{\cite{Ran90} got $\alpha=86^o$}.  Assuming the 
radiation are from single pole, we find that the phase resolved lower bounds are 
roughly around 250 km for various inclination angles,  which is concord with our 
estimation in \S\ref{secest}.

\begin{figure*} \center \includegraphics[height=4.0 in]{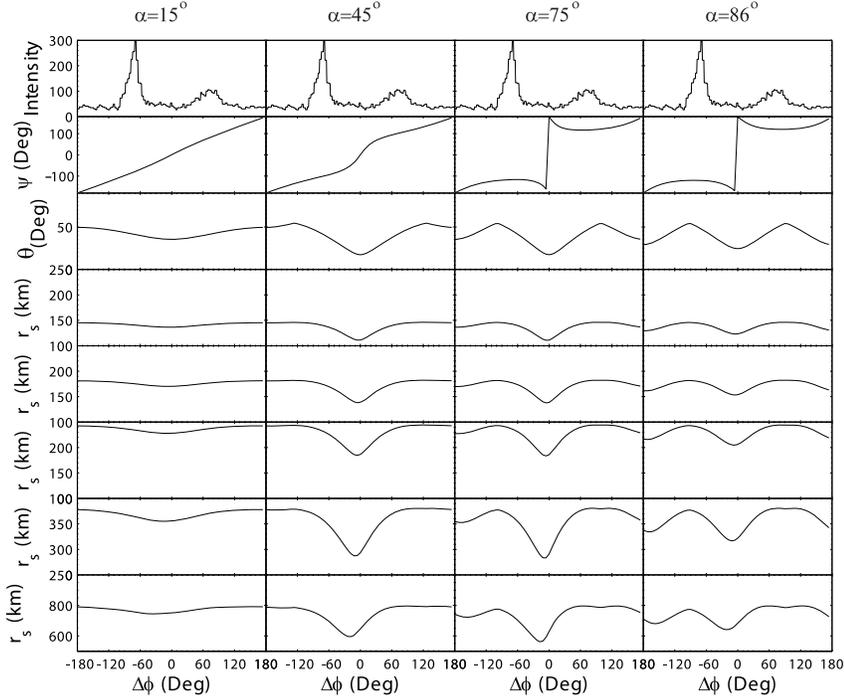}
	\caption{The altitude lower bounds for different pulse longitudes. Each 
	column corresponds to results for different inclination angle $\alpha$, as 
	labeled on the top. The x-axes are the pulse longitudinal phase. The first 
	row is the pulse profile of Crab observed in above 100 MeV band, the data 
	comes from \citep{Thom01} and is not very different from recent data from 
	FERMI \citep{F09b}; the second and the third row are the azimuthal angle and 
	polar angle of radiation source in magnetic polar coordinate, respectively; 
	the fourth to the eighth row are for the $\tau_{\rm th}=1$ radiation 
	altitude lower bounds for $\varepsilon=5$ GeV,
	$\varepsilon=10$ GeV,
	$\varepsilon=25$ GeV,
	$\varepsilon=100$ GeV, and $\varepsilon=1$ TeV, respectively.  During the 
	calculation, we take $B_0=3.7\times 10^{12}$ Gauss \citep{MHT05}, 
	$\zeta=60^{o}$ \citep{WHT00, Wang03}, and $p=0.033$\citep{MHT05} s as
	Crab pulsar parameters.}
\label{fig:phra}
\end{figure*}

For pulsar with inclination angle $\alpha$ and the view angle $\zeta$, we can 
use Eq.~\ref{rho}, \ref{eqtheta}, and \ref{eqvarphi} to calculate the magnetic 
polar angle $\theta_{\rm s}$ and magnetic azimuthal angle $\phi_{\rm s}$ for 
radiation source, which contribute to pulse profile with longitude $\Delta\Phi$.  
The we use bi-section method to solve $\tau(r_{\rm tr})=\tau_{\rm th}$ respected 
to $r_{\rm tr}$ to determine the altitude lower bounds, below which the photon 
escape probability is less than $e^{-\tau_{\rm th}}$.  The numerical results are 
given in Fig.~\ref{fig:phra} for $\tau_{\rm th}=1$ case. 

\section{Discussions and conclusions}
\label{seccon}
In this paper, we derive the basic formalism for calculating the effect of 
$\gamma$-B absorption for $\gamma$-ray photons in pulsar magnetosphere. We have
considered magnetic field line bending, aberration effect, rotational effects, 
gravitational red-shift of photons and gravitational effect on magnetic field 
intensity. 

In particular, we calculate the altitudes for last trapping surface, at which 
the photon generated is optical thick ($\tau=1$) due to $\gamma$-B processes 
both analytically (\S~\ref{secest})  and numerically (\S\ref{secdet}).
We regard such altitudes as the lower bounds for $\gamma$-ray radiation 
altitude. The analytical estimation and numerical results get good agreement.  
The numerical calculation suggest slightly higher value (20\%) for $r_{\rm tr}$ 
than analytical calculation, due to approximation for analytical method. 

The altitude lower bounds are calculated for different longitudes. For Crab 
pulsar, the altitude lower bound is about 100 km, 150 km,250 km, 400 km, and 800 
km, for 5 GeV, 10 GeV, 25 GeV, 100 GeV, and 1 TeV$\gamma$-ray radiation 
respectively for various inclination angle. Note that the $r_{\rm tr}$ is larger
around $\Delta \phi=0$. This is a geometrical effect.
Because the curvature radius of magnetic field lines are smaller around $\Delta 
\phi=0$, the transverse magnetic field intensity $B_{\rm \bot}$ and the 
absorption coefficient $\kappa$ become smaller.  Although beyond the measured 
cut-off energy, 10 GeV pulsed emission is confirmed, 150 km radiation altitude 
lower bounds could falsify the classical polar cap
model for puslar's $\gamma$-ray radiation \citep{DH94}.  However it can not 
discriminate present pulsar $\gamma$-ray radiation model including
slop gap model \citep{HSD08}, outer gap model \citep{CHR86a, CHR86a, CR94, 
CRZ00} and inner annular gap
model \citep{QLWXH03}. We can neither exclude possibility for a high altitude 
model producing a low energy cut-off as indicated in \citep{HSD08}, since
such cut-off may be due to intrinsic physical conditions. 

We see that the lower bounds $r_{\rm tr}$ is not sensitive to $\tau_{\rm th}$,  
mainly due to the logarithmic dependence of $\tau_{\rm th}$, as we have already 
seen in Eq.~\ref{eq:eqtr}. Thus the difference between $r_{\rm tr}$ for 
$\tau_{\rm th}=1$ and $\tau_{\rm th}=10$ is less than 10\%.

The magnetic field components of higher order magnetic momentum decrease much 
faster than
that of the dipole magnetic momentum. Therefor for the emission from high 
altitudes ($r/R\sim 10^{-2}$),
we can thus ignore the higher order magnetic momentum than dipole, although the 
higher order magnetic momentum are important near the star
surface \citep{RS75}.

Our results give different value compared to \cite{Bar04}. The main reason is 
the difference of treating photon characteristic absorption length. A brief 
comparison is given in the appendix of this paper. It is shown that 
\cite{Bar04}'s result under-estimates the characteristic absorption 
length, such that it needs stronger magnetic field to achieve absorption, which 
under estimate the $r_{\rm tr}$. 

The cut-off is not a good radiation location indicator. Firstly, it is not only 
sensitive to data quality but also sensitive to the data reduction processes, 
i.e. how the cut-off is fitted \citep{MAG09}. Secondly, there are still pulsed 
emission photons coming from the cut-off tails. Thirdly, one needs to model the 
initial photon population to make reasonable result for cut-off energy. Thus we 
only use the observation fact that photons of energy $\varepsilon$ are detected, 
then we discuss where is the lowest possible position (i.e. $r_{\rm tr}$), above 
which the photons can just propagate freely without strong absorption by pulsar 
magnetic field.  Such $r_{\rm tr}$ is well defined. Future measurement of 
cut-off energy  will
not change the conclusion here, as far as 10 GeV pulse emission of the Crab 
pulsar has been confirmed, the radiation altitude of 150~km we calculated for 
the lower bounds holds.

Seven pulsars were detected in $\gamma$-ray bands previously \citep{Thom08}, and 
more than 40 $\gamma$-ray pulsars are just found recently
e.g. \citep{F09a, PP09}. FERMI \citep{FERMI09} and AGILE are now seeking for 
more $\gamma$-ray pulsars. If young $\gamma$-ray pulsars with stronger magnetic 
field are observed at higher energy (e.g. 1 Tev) in the future, there would be a 
great challenge to the slop gap model and annular gap model.
Future $\gamma$-ray pulsar searching and the follow up observations using ground 
based Cherenkov telescope ares expected to test present $\gamma$-ray radiation 
models for pulsars. 

We are grateful to C.K. Chou for reading the paper. We are also grateful
to L. Guillemot and A. Jessner for reading the paper and giving valuable
comments improving the paper. This work was supported by NSFC (10833003,
10778611, 10821061) and the National Basic Research Program of China
(Grant 2009CB824800).

\appendix
\section{Comparison between Baring's results and results in this paper}

This paper answers a different question compared with \cite{Bar04}.  We are 
discussing the altitude lower bound for observed photons with specific energy, 
while \cite{Bar04} discussed the spectra cut-off for collective photons.  
Nevertheless it is interesting to see the difference of the results in 
\cite{Bar04}'s and here. It turns out that we will get \cite{Bar04}'s results, 
if the characteristic absorption length $\lambda_{\rm c}$ is taken to be 
constant, which corresponds to the case where radiation region is near the star 
surface. 

Following Eq.~19 of Ruderman and Sutherland (1975), the $\gamma-B$
absorption criteria is $\varepsilon_{\rm MeV} B_{\bot}/B_{\rm cr}>\chi$, where 
Such criteria is also adopted by \cite{Bar04}.
Due to the curvature of magnetic field, we have $B_{\bot}\sim B \sin\theta_{\rm 
i}\sim B\lambda_{\rm c}/ r_{\rm cur} \sim B_{0} \frac{R^3}{r_{\rm 
s}^3}\frac{\lambda_{\rm c}}{r_{\rm cur}}$.
So the absorption criteria is

\begin{equation}
	\varepsilon_{\rm MeV} \frac{B_0}{B_{\rm cr}}\left(\frac{R_0}{r_{\rm 
	s}}\right)^3 \frac{\lambda_{\rm c}}{r_{\rm cur}} > \chi
\end{equation}

Substitute $r_{\rm cur}=\frac{4}{3} \left(\frac{p r_{\rm s} c}{2
\pi}\right)^{1/2}$, we have \begin{equation}
	\varepsilon_{\rm Max, [MeV]}<\chi \frac{B_{\rm
	cr}}{B_0} \left(\frac{r_{\rm s}}{R_0}\right)^3 \frac{(p r_{\rm s} c/2
	\pi)^{1/2}}{\lambda_{\rm c}},
\end{equation} which is just

\begin{equation}
	\varepsilon_{\rm Max, [GeV]}=\left[10^{-2} \chi \sqrt{c/2\pi} 
	\frac{R_{0}^{1/2}}{\lambda_{\rm c}} \right ]\sqrt{p} \left(\frac{r_{\rm 
	s}}{R_0} \right )^{1/2}
	\frac{0.1 B_{\rm cr}}{B_0} \left(\frac{r_{\rm s}}{R_0}\right)^3 ,
\end{equation}

If we take $\lambda_{\rm c}/\chi=17$, we just get \cite{Bar04}'s results
\begin{equation}
	\varepsilon\simeq0.4 \sqrt{p}\left(\frac{r_{\rm s}}{R_{0}}\right)^{1/2} {\rm 
	max} \left\{1, 0.1 \frac{B_{\rm cr}}{B_0} \left(\frac{r_{\rm 
	s}}{R_0}\right)^3 \right\} {\rm GeV}.
	\label{B04}
\end{equation}

\begin{figure*} \center \includegraphics[height=1.5in]{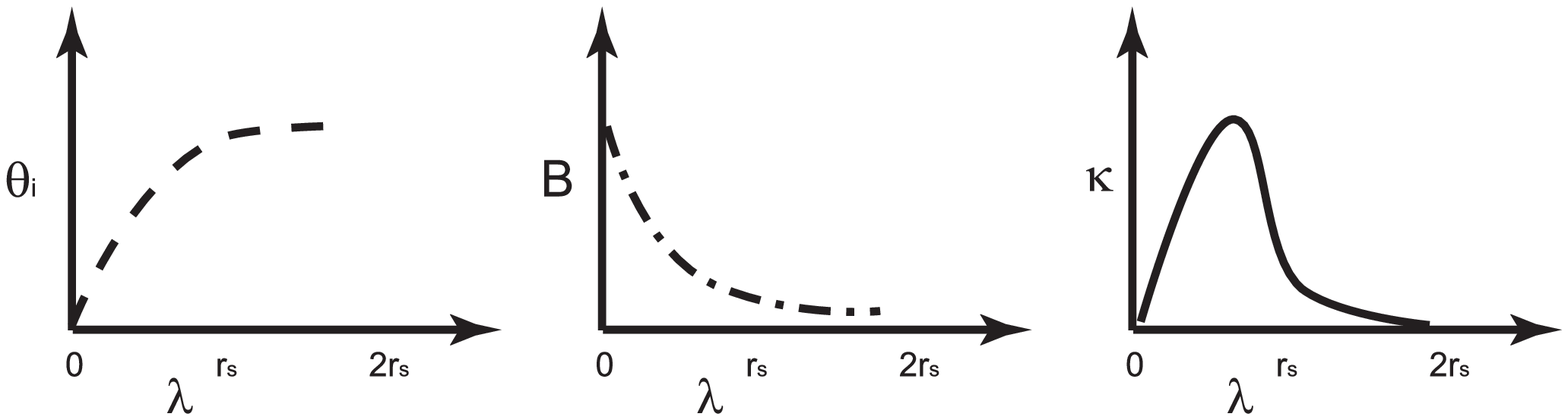}
	\caption{The illustration for the relation between $\gamma$-B impact angle 
	$\theta_{\rm i}$, magnetic field intensity $B$, absorption coefficient 
	$\kappa$ and photon propagating length $\lambda$. $\theta_{i}, B$ and 
	$\kappa$ follow the dashed, dot-dashed and solid curve respectively when 
	$\lambda$ grows from 0 to large values.}
\label{fig:ktb} \end{figure*}

If we take $\lambda_{\rm c}=r_{\rm s}$, we go back to Eq.~\ref{eq:eqtr}.  The 
reason for making such choice is illustrated in Fig.~\ref{fig:ktb}. There are 
two comparative factors dominant the tendency of absorption coefficient 
$\kappa$.  Firstly, the impact angle $\theta_{i}$ between photon direction and 
local magnetic field grows with $\lambda$ then saturated to a limited value 
due to pure geometrical relation.  Secondly, magnetic field intensity 
decrease approximately following $(r_{\rm s}+\lambda)^{-3}$ due to dipole field 
configuration.  Thus the perpendicular magnetic field strength $B_{\bot}\sim 
B\theta_{\rm i}$ follows relation $\lambda/(r_{\rm s}+\lambda)^{-3}$.  The 
absorption coefficient $\kappa$ depends on $B_{\bot}$ exponentially. So 
$\theta_{i}, B$ and $\kappa$ follow the dashed, dot-dashed and solid curve in 
Fig.~\ref{fig:ktb} respectively.  Clearly the effective absorption length takes 
value of $r$.  Precisely speaking, there must be some numerical fact $\eta$ (due 
to integration and geometry), where $\eta \simeq 1$,  such that $\lambda_{\rm 
c}=\eta r_{\rm s}$.  In practical, this $\eta$ can be regarded as been accounted 
and put into the threshold optical depth $\tau_{\rm th}$.  Due to the dependence 
of $r_{\rm tr}$ on $\tau_{\rm th}$ is logarithmic (see Eq.~\ref{eq:eqtr}), such 
correction of $\eta$ is not important. This is why the analytical estimation is 
concord with the numerical calculation developed in section \ref{secdet}. 

Clearly at high altitude, the characteristic absorption length $\lambda_{\rm c}$ 
is no long a constant and must be treated as a function of radiation altitude.  
Because of the implicit assumption of a constant characteristic absorption 
length, \cite{Bar04} results under-estimate the characteristic absorption length 
$\lambda_{\rm c}$ so does the optical depth. This leads to smaller $r_{\rm tr}$ 
as we have seen for high altitude radiation. We have used $\lambda_{\rm c} \sim 
r_{\rm s}$ as argued in this paper to derive Eq.~\ref{eq:eqtr}, which is later 
checked by our numerical results.  For low energy photons (a few GeV), our 
results agree with \cite{BH01} because the characteristic absorption length is 
about the size of star.  

\bsp
\label{lastpage}
\end{document}